\documentclass[sigconf,screen]{acmart}
\usepackage{balance}
\usepackage[utf8]{inputenc}
\usepackage[russian,english]{babel}
\usepackage{ragged2e}
\usepackage{lineno}

\AtBeginDocument{%
  \providecommand\BibTeX{{%
    \normalfont B\kern-0.5em{\scshape i\kern-0.25em b}\kern-0.8em\TeX}}}

\begin{document}
\author{Francesco Pierri}
\affiliation{%
  \institution{Information Sciences Institute, University of Southern California, Los Angeles, USA}
    \country{}
  }
\affiliation{%
  \institution{Politecnico di Milano, Milano, Italy}
    \country{}
  }
\author{Luca Luceri}
\affiliation{%
  \institution{Information Sciences Institute, University of Southern California, Los Angeles, USA}
    \country{}
  }
\author{Nikhil Jindal}
\affiliation{%
  \institution{Information Sciences Institute, University of Southern California, Los Angeles, USA}
  \country{}
  }
\author{Emilio Ferrara}
\affiliation{%
  \institution{Information Sciences Institute, University of Southern California, Los Angeles, USA}
    \country{}
  }
\title{Propaganda and Misinformation on Facebook and Twitter \\ during the Russian Invasion of Ukraine}

\begin{abstract}
Online social media represent an oftentimes unique source of information, and having access to reliable and unbiased content is crucial, especially during crises and contentious events. We study the spread of propaganda and misinformation that circulated on Facebook and Twitter during the first few months of the Russia-Ukraine conflict. By leveraging two large datasets of millions of social media posts, we estimate the prevalence of Russian propaganda and low-credibility content on the two platforms, describing temporal patterns and highlighting the disproportionate role played by superspreaders in amplifying unreliable content. We infer the political leaning of Facebook pages and Twitter users sharing propaganda and misinformation, and observe they tend to be more right-leaning than the average. By estimating the amount of content moderated by the two platforms, we show that only about 8-15\% of the posts and tweets sharing links to Russian propaganda or untrustworthy sources were removed. Overall, our findings show that Facebook and Twitter are still vulnerable to abuse, especially during crises: we highlight the need to urgently address this issue to preserve the integrity of online conversations.
\end{abstract}




\maketitle

\section{Introduction}
Eight years after the annexation of Crimea, on February 24, 2022, Russia invaded Ukraine, with unprecedented consequences for the rest of the world.\footnote{\url{https://en.wikipedia.org/wiki/2022_Russian_invasion_of_Ukraine}} The ongoing conflict has led to a global energy crisis and food shortages, pushing millions of Ukrainian citizens to flee from the country as refugees.\footnote{\url{https://www.cbsnews.com/news/ukraine-russia-death-toll-invasion/}} Media outlets and news agencies around the world started reporting about the conflict from strikingly different points of view.\footnote{\url{https://www.theatlantic.com/technology/archive/2022/03/russia-ukraine-war-propaganda/626975/}} For instance, the Western press immediately condemned the invasion, whereas Russia justified its ``special operation'' as a mission to remove alleged Nazis from Ukraine.\footnote{\url{https://news.un.org/en/story/2022/09/1127881}} Besides, other countries such as China and India blamed NATO's expansion for causing the war and, at the same time, advocated for diplomacy \cite{hanley2022special}.

% Online social media represent a vital and oftentimes unique source of information for individuals around the world \cite{tang2013facebook,chan2016social}. In recent times, an ever-increasing concern has been raised in relation to information disorders and coordinated harm that take place on social platforms \cite{ferrara2016rise,Lazer-fake-news-2018}. These cyber-social threats are particularly relevant during crises, when access to accurate and reliable information is crucial \cite{gallotti2020assessing}. In parallel with the military invasion of Ukraine, Russia actively engaged in promoting propaganda and mis/disinformation about the war, with the goal of manipulating public opinion to undermine support for Ukraine \cite{alyukov2022propaganda}. Russian meddling with the democratic processes of other countries has been already reported in the past. The 2016 US Presidential election represents a prime example of Russian interference campaign on social media \cite{badawy2018analyzing,carroll2017st,popken2018twitter} --- recently described in the Mueller report \cite{mueller2019mueller} as a ``sweeping and systematic'' attack to the US election system. In this orchestrated campaign, Russia's ``Internet Research Agency" (IRA) employed bots (i.e., software-controlled accounts) \cite{bessi2016social,howard2018algorithms} and trolls (i.e., state-backed human agents) \cite{badawy2019characterizing,luceri2020detecting} to sow discord, spread misinformation, and diffuse politically biased content.

In parallel with the military invasion of Ukraine, Russia actively engaged in promoting propaganda and mis/disinformation about the war, with the goal of manipulating public opinion to undermine support for Ukraine \cite{alyukov2022propaganda}. Russian meddling with other countries' democratic processes has been extensively documented. The 2016 U.S. Presidential election represents a prime example of Russian interference on social media \cite{badawy2018analyzing,carroll2017st,popken2018twitter} ---  described in the Mueller report \cite{mueller2019mueller} as a ``sweeping and systematic'' attack on the U.S. democracy. In that orchestrated campaign, Russia's ``Internet Research Agency'' (IRA) employed bots (i.e., software-controlled accounts) \cite{bessi2016social,howard2018algorithms} and trolls (i.e., state-backed human agents) \cite{badawy2018analyzing,badawy2019characterizing,luceri2020detecting} to sow discord~\cite{stella2018bots}, spread misinformation~\cite{chen2021covid, chen2022charting, sharma2022characterizing}, ignite conspiracy theories~\cite{ferrara2020types, muric2021covid, sharma2022characterizing, wang2023identifying}, and diffuse politically biased content online~\cite{jiang2020political, jiang2021social, rao2021political, majo2021role, jiang2023retweet}. Ever-increasing concerns are continuously raised in relation to information disorder and coordinated harm that take place on social platforms \cite{ferrara2016rise,Lazer-fake-news-2018,sharma2021identifying,ferrara2022twitter}. These cyber-social threats are particularly relevant during crises, when access to accurate and reliable information is crucial \cite{gallotti2020assessing}.

\subsection{Research Questions \& Contributions}
In this paper, we provide a longitudinal study of the spread of misinformation and propaganda about the ongoing Russian invasion of Ukraine on two mainstream social platforms -- Facebook and Twitter -- over a period of 4 months. To this end, we leverage a large-scale data collection of almost 20M Facebook posts, which generated over 2.9 billion interactions, and more than 250M tweets, in order to track and assess the prevalence of news articles originating from Russian-state outlets and low-credibility news websites, compared to high-credibility content shared by a representative sample of reputable sources. 

We formulate and address the following research questions:
\begin{itemize}
\item[\textbf{RQ1}:] \emph{What is the prevalence of low-credibility content and Russian propaganda  on Facebook and Twitter during Russia's invasion of Ukraine?}
\item[\textbf{RQ2}:] \emph{Who are the superspreaders of Russian propaganda and low-credibility content?}
\item[\textbf{RQ3}:] \emph{What is the inferred political leaning of accounts sharing Russian propaganda and low-credibility content?}
\item[\textbf{RQ4}:] \emph{What amount of Russian propaganda and low-credibility content is removed by the platforms?}
\end{itemize}
We provide a number of contributions in addressing these research questions. We show that Russian propaganda becomes less prevalent after the invasion, following platforms' intervention, European sanctions on state outlets and Russian ban on Facebook and Twitter, but it does not disappear completely. Low-credibility content, on the other hand, exhibits a stable trend in the number of reshares and retweets throughout the period of analysis. 
We highlight the role played by a certain group of influential and verified Facebook pages and Twitter users, showing that a handful of them accounts for 60-80\% of all the reshares and retweets of problematic content. We infer the political leaning of accounts sharing Russian propaganda and low-credibility content, finding that they skew toward the right end of the political spectrum compared to the average account. Finally, we estimate the amount of problematic content moderated by the two platforms, finding that only 8-15\% of posts and tweets sharing links to Russian propaganda and low-credibility content were actually removed. Our findings add to extant literature that aims to shed light on the information disorder taking place on online social platforms, especially during global crises, and advocate for further interventions on this matter in order to preserve the integrity of democratic processes.

% The outline of the paper is the following: in the next section we outline previous contributions related to the present work; then, we describe the methods employed to collect the data and perform our analyses; next, we provide results to address each of the research questions formulated above. Finally, we draw conclusions, pinpoint limitations to our work and suggest directions for future research.

\section{Related Work}
In the following, we review existing contributions that tackle information disorders in the specific context of the Russian invasion of Ukraine. We refer the reader to \cite{pierri2019false,zhou2020survey,ruffo2023studying} for a broader overview of the literature on misinformation, disinformation, and other forms of cyber social threats in online social media.

The earliest investigations of suspicious activity on Twitter following the Russian invasion of Ukraine date back to mid-March 2022 \cite{chen2022tweets,osomewp1,osomewp2}. Different groups noted peaks in the creation of new accounts around the day of the invasion (February 24, 2022), and revealed the presence of coordinated groups of users spamming and boosting hate speech. By means of a qualitative analysis, however, they showed that most of the related messages shared on Twitter during the early phase of the conflict were genuine or benign, with pro-Ukraine messages being much more prevalent than pro-Russia ones.

\citet{caprolu2022characterizing} used a mixed-methods approach to analyze 5M+ tweets related to the conflict, showing little evidence of disinformation campaigns on Twitter, contrary to mainstream reports.

\citet{park2022voynaslov} introduced the \textit{VoynaSlov} dataset to help researchers study information manipulation campaigns at play on Twitter and VKontakte during the conflict. The collection contains over 38M posts from Russian media outlets shared on the two platforms, which were examined by the researchers to investigate agenda-setting and framing effects.

\citet{hanley2022happenstance} provided two different contributions on this topic. In the first, they used sentence-level topic analyses to study Russian propaganda on Reddit shared between January and April 2022, finding that approximately 40\% of the comments in the \textit{r/Russia} subreddit promote Russian mis/disinformation. In the second \cite{hanley2022special}, they used a combination of sentiment and topic analysis to study western, Russian, and Chinese media on Twitter and Weibo. They found that, while the focus of the western press was on military and humanitarian aspects of the war, Russian media focused on justifying their ``special military operation'', whereas Chinese news insisted on the conflict’s diplomatic and economic consequences in the geopolitical landscape.

\citet{pierri2022does} analyzed Twitter account moderation efforts during the first months of the conflict, by identifying peaks of suspicious account creation and suspension, and characterized behaviours that more frequently lead to account suspension. They showed that many accounts got suspended a few days after their creation, most likely because they made excessive use of replies, spam and harmful messages. 

Finally, \citet{smart2022istandwithputin} and \citet{geissler2022russian} studied the activity of automated accounts sharing pro- and anti-Russia hashtags on Twitter, in order to quantify how bots influence human accounts in online conversations as well as highlight their role in spreading Russian propaganda and disinformation.

\section{Methods}
\subsection{Data collection}
\begin{figure}[!t]
    \centering
    \includegraphics[width=\linewidth]{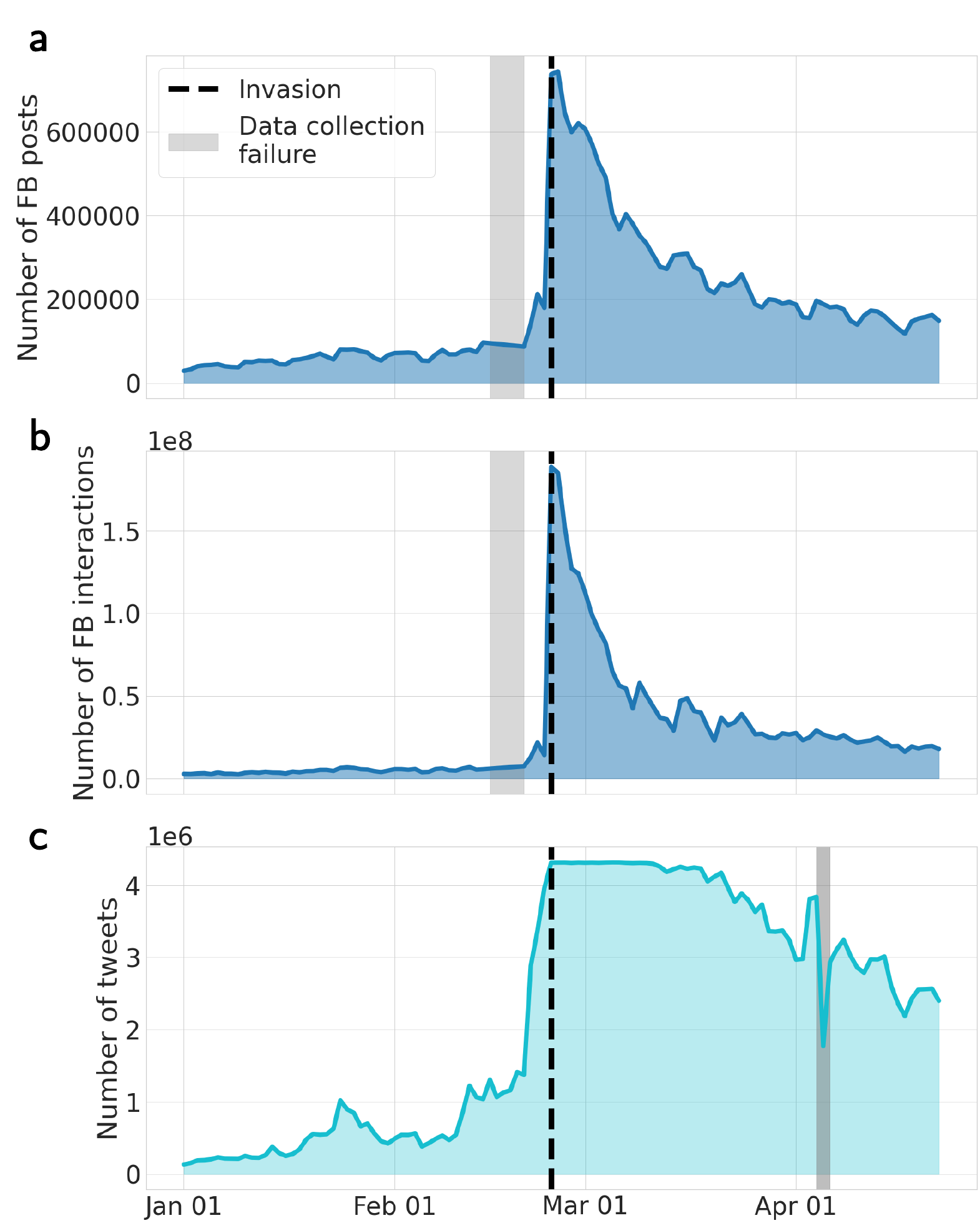}
    \caption{Time series of the daily number of Facebook posts (a), the interactions they generated (b), and tweets (c) in our dataset. The vertical dashed line indicates the day of the invasion (February 24, 2022). Grey shades indicate data collection failures. Specifically, for Facebook data they were due to API malfunctioning whereas for Twitter data they were due to network issues.}
    \label{fig:general-ts}
\end{figure}

In our analyses, we leverage two distinct datasets of Facebook and Twitter social media posts shared over a period of 4 months (January 1, 2022 - April 24, 2022).
We collected Facebook data by employing CrowdTangle, a public tool owned and operated by Meta~\cite{crowdtangle} that allows searching the entire collection of Facebook posts shared by public pages and groups that have a certain amount of followers or that were added by other researchers on the platform.\footnote{See the official documentation for more details on the coverage: \url{https://help.crowdtangle.com/en/articles/1140930-what-data-is-crowdtangle-tracking}} Throughout the text, we will refer to them as `accounts' for consistency with Twitter, even though they do not represent individual users. Crowdtangle API does not allow to collect data in a streaming fashion, therefore we queried the \texttt{/posts/search} endpoint\footnote{\url{github.com/CrowdTangle/API/wiki/Search}} weekly,\footnote{In particular, we used a sliding time window.} using a set of over 40 keywords (in English, Russian and Ukrainian language) related to the conflict and introduced in \cite{osomewp1,osomewp2}. The resulting dataset contains 19.5M posts, shared by 1.1M unique pages and groups, generating over 2.9 billion interactions (shares, comments, reactions, etc). 
% A sample of keywords is available in Table \ref{tab:fb-keywords}. 
We show the daily number of posts and interactions in panels \textbf{a} and \textbf{b} of Figure \ref{fig:general-ts}. We provide access to the IDs of these posts, which can be used to retrieve the data by means of Crowdtangle, in the repository associated with this paper.\footnote{\url{github.com/frapierri/uk-ru_propaganda_misinformation_tw_fb}}

% \begin{otherlanguage*}{russian}
% \begin{table}[!t]
% \begin{tabular}{llll}
% ukrain & zelensky & nato & donezk \\ \hline
% украина & nord stream & donetsk & київ \\ \hline
% kiev & ukrainets & russland & україн \\ \hline
% putin & ukrayiny & kremlin & зеленський \\ \hline
% \end{tabular}
% \caption{Sample of keywords used to collect Facebook data.}
% \label{tab:fb-keywords}
% \end{table}
% \end{otherlanguage*}

We combined two data sources for Twitter: for the period February 22, 2022 - April 24, 2022 we referred to an existing dataset \cite{chen2022tweets} collected through the \texttt{Standard v1.1 Streaming} endpoint\footnote{\url{https://developer.twitter.com/en/docs/twitter-api/v1}} that contains tweets matching over 30 keywords (in English, Russian and Ukrainian language) related to Russia's invasion of Ukraine. These were identified with a snowball sampling approach by looking at trending topics and hashtags, and they largely overlap with those specified in the Facebook data collection. We further employed, in May 2022, the historical Search API v2\footnote{https://developer.twitter.com/en/docs/twitter-api/tweets/search/introduction} to collect tweets in the period January 1, 2022 - February 21, 2022, using the same set of keywords. 
% A sample is available in Table \ref{tab:tw-keywords}, and the full list is available in the repository associated with the dataset\footnote{Link omitted for blind review}. 
The repository associated with the dataset paper \cite{chen2022tweets} contains tweet IDs that can be re-hydrated by querying the Twitter API\footnote{\url{github.com/echen102/ukraine-russia}}. Overall, the dataset contains almost 250M tweets shared by 15M unique users. We remark that the streaming endpoint filters tweets that match a defined query in a real-time fashion up to 1\% of the global stream \cite{morstatter2013sample}. As it can be seen in panel \textbf{c} of Figure \ref{fig:general-ts}, we likely hit the rate limit during the weeks following the invasion (February 24, 2022), when the data volume caps at around 4M daily tweets.

In terms of language distribution, we observe that in Facebook data most posts do not have a defined language (42\%), according to the \texttt{languageCode} parameter provided by Crowdtangle, whereas over 16\% of the posts are shared in English, 8\% in Ukrainian and 3\% in Russian. In Twitter data, based on the \texttt{lang} parameter provided by the API, the majority of posts are shared in English (over 70\%), with less than 2\% of posts in Ukrainian and Russian or with undefined language.

% \begin{otherlanguage*}{russian}
% \begin{table}[!t]
% \begin{tabular}{llll}
% ukraine  & zelensky & nato & donezk \\ \hline
% russia & minsk & moscow & SlavaUkraini \\ \hline
% kiev & fsb & russland & фсб \\ \hline
% putin & kgb & Россия & luhansk \\ \hline
% \end{tabular}
% \caption{Sample of keywords used to collect Twitter data.}
% \label{tab:tw-keywords}
% \end{table}
% \end{otherlanguage*}

\subsection{Labeling sources of online information}
To identify reliable and unreliable content shared on Facebook and Twitter, we compiled three lists of news websites corresponding respectively to Russian propaganda, low-credibility and high-credibility news outlets. We follow a distant-supervision approach, widely adopted in the literature \cite{shao2018spread,grinberg2019fake,bovet2019influence}, to label news articles based on the reliability of the source. 

We referred to the \textit{VoynaSlov} dataset \cite{park2022voynaslov} to obtain a list of 23 state-affiliated Russian media websites, manually verified by a fluent Russian speaker, which have been flagged for sharing unsubstantiated claims and Russian propaganda about the war. We also added \texttt{yandex.ru}, the top Russian search engine that has been repeatedly reported for indexing and promoting propaganda websites. A sample of websites are available in Table \ref{tab:ru-websites}, whereas the full list is available in the repository associated with this paper.

\begin{table}[!t]
    \centering
\begin{tabular}{lll}
sputniknews.com &
rt.com &
redfish.media \\ \hline
tass.com &
tass.go & go.tass.ru \\ \hline
ria.ru &
ruptly.tv &
m24.ru \\ \hline
\end{tabular}
    \caption{Sample of Russian propaganda websites.}
    \label{tab:ru-websites}
\end{table}

For what concerns low-credibility news websites, we referred to the Iffy Index of Unreliable Sources,\footnote{\url{https://iffy.news/index/}} a list of over 600 low-credibility domains based on information provided by the Media Bias/Fact Check website (MBFC, \url{mediabiasfactcheck.com}) in which political leaning is not a factor. Throughout the text and for sake of simplicity, we will use \textit{misinformation} to refer to both Russian propaganda and low-credibility news, although oftentimes this term is used to refer to false or incorrect information that is shared unintentionally as opposed to \textit{disinformation}, which is instead produced to deliberately deceive readers.

Finally, we refer directly to MBFC to gather a representative sample of reputable news outlets as a reference for high-credibility content. In particular, we picked 10 websites from the most shared ones on both platforms, among those with a high level of ``Factual Reporting'' and ``Credibility Rating'' and spanning the entire political spectrum. The list of websites is available in Table \ref{tab:high-websites}.

\begin{table}[!t]
    \centering
\begin{tabular}{lll}
nytimes.com & reuters.com & wsj.com \\ \hline
nbcnews.com & washingtonpost.com & ft.com \\ \hline
businessinsider.com & apnews.com & bloomberg.com \\ \hline
bbc.com &  &  \\ \hline
\end{tabular}
    \caption{List of high-credibility news websites.}
    \label{tab:high-websites}
\end{table}

\subsection{Inferring political leaning}
To infer political bias of Facebook and Twitter accounts, we consider the liberal-conservative spectrum defined by a score between $-1$ (liberal) and $+1$ (conservative). We assign a political alignment score to each post containing a URL, and then we average at the user level to measure accounts' political alignment. Specifically, we leverage the political bias annotations of over 19K websites provided by \citet{chen2021neutral}, whose scores were obtained from the sharing activity of Twitter accounts associated with registered U.S. voters \cite{robertson2018auditing}. Similarly to \citet{chen2021neutral}, we did not consider links to social media platforms such as Twitter, Facebook, Instagram, YouTube. 

\begin{figure*}[!t]
    \centering
 \includegraphics[width=\linewidth]{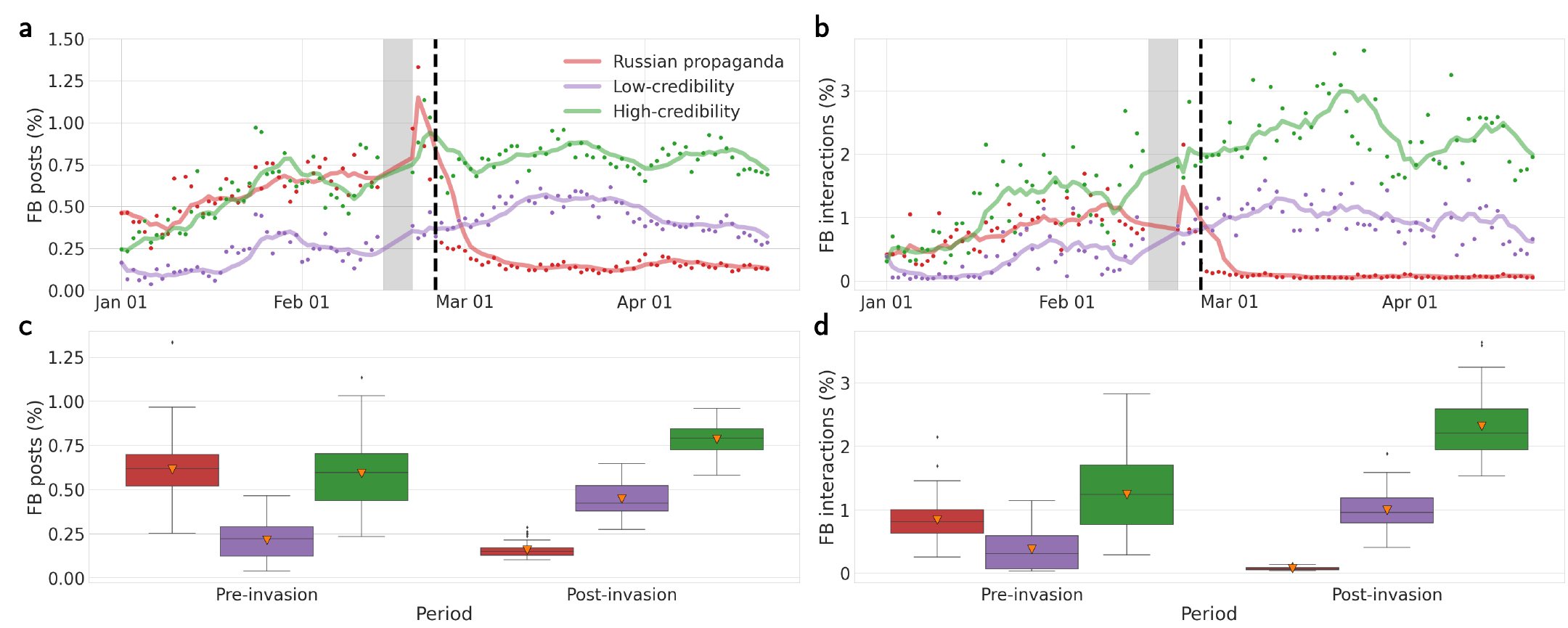}
    \caption{Time series and distributions of the daily proportion of Facebook posts \textbf{(a,c)} and the interactions \textbf{(b,d)} they generated when sharing links to Russian propaganda, low-credibility and high-credibility news websites with respect to all the posts in our dataset. Solid lines represent 7-day rolling averages. The dashed vertical line indicates the day of the invasion (February 24, 2022), and the grey shade indicates a data collection failure. Median values in \textbf{(c)} are: Russian Propaganda (Pre-invasion = 0.62\%, Post-invasion = 0.15\%); Low-credibility (Pre-invasion = 0.22\%, Post-invasion = 0.42\%); High-credibility (Pre-invasion = 0.59\%, Post-invasion = 0.79\%). Median values in panel \textbf{(d}) are: Russian propaganda (Pre-invasion = 0.81\%, Post-invasion = 0.06\%); Low-credibility (Pre-invasion = 0.30\%, Post invasion = 0.95\%); High-credibility (Pre-invasion = 1.24\%, Post-invasion = 2.20\%). Triangles in (c) and (d) represent the mean value of the distribution.}
    \label{fig:prevalence-fb}
\end{figure*}

\subsection{Identifying removed content}
To identify content removed by Facebook and Twitter, we carried out two different retrieval procedures, both on October 15th, 2022.
For what concerns Facebook, we queried the \texttt{/post/:id} endpoint\footnote{\url{https://github.com/CrowdTangle/API/wiki/Posts}} to retrieve all posts already collected in our dataset. However, since the API imposes severe limitations on the number of calls that can be made ($\sim$6 calls per minute), we only retrieved four random samples of 10K posts, one for each kind of content (those containing Russian propaganda, those sharing low-credibility and high-credibility content, and a set of random posts). We therefore labeled as `removed' those posts that were not returned by the API ($\sim$3K out of 40K posts). Some details on Facebook moderation policies are available on their website.\footnote{\url{https://transparency.fb.com/it-it/policies/community-standards/}}

We identified removed tweets using the \texttt{compliance/jobs} endpoint via \texttt{twarc2}.\footnote{\url{https://twarc-project.readthedocs.io/en/latest/twarc2_en_us/\#compliance-job}} Specifically, we queried Twitter for all the tweets collected using the streaming endpoint, obtaining almost 40M tweets that were removed by the platform either because they violated the Terms of Service rules, or the author's account was suspended/deleted. The reason for not including tweets collected with the historical search endpoint is that this collection, which was performed a few months after the beginning of the conflict, obviously does not include tweets that were removed before the collection process. However, this does not affect our results since well over 90\% of the tweets in the overall dataset were collected through the streaming endpoint.
More details about reasons for suspension are available in the Twitter documentation.\footnote{\url{https://help.twitter.com/en/managing-your-account/suspended-twitter-accounts}}

\section{Results}
\subsection{Prevalence of misinformation about the conflict}
In this section, we compare the prevalence of social media posts sharing links to Russian propaganda and low-credibility content with respect to more reputable sources. 
On the one hand, we remark that we only consider a handful of representative sources of high-credibility content. On the other hand, we only identify misinformation shared via news articles, and we are not tracking unsubstantiated claims that are shared in messages, images, and videos. Thus, in both cases, their prevalence should be seen as a lower-bound estimate.

On Facebook, we identified 51,269 posts (0.25\% of all posts) sharing links to Russian propaganda outlets, generating 5,065,983 interactions (0.17\% of all interactions); 80,066 posts (0.4\% of all posts) sharing links to low-credibility news websites, generating 28,334,900 interactions (0.95\% of all interactions); and 147,841 posts sharing links to high-credibility news websites (0.73\% of all posts), generating 63,837,701 interactions (2.13\% of all interactions). As shown in Figure \ref{fig:prevalence-fb}, we notice that the number of posts sharing Russian propaganda and low-credibility news exhibits an increasing trend (Mann-Kendall $P < .001$), whereas after the invasion of Ukraine both time series yield a significant decreasing trend (more prominent in the case of Russian propaganda); high-credibility content also exhibits an increasing trend in the Pre-invasion period (Mann-Kendall $P < .001$), which becomes stable (no trend) in the period afterward. These patterns are shown in panel \textbf{(a)}. Interestingly, the number of posts sharing Russian propaganda is higher than low-credibility sources on an average day, in the Pre-invasion period (two-way Mann-Whitney $P < .001$). However, the number of posts sharing links to Russian state outlets considerably drops after the invasion of Ukraine, due to Facebook's policies that regulated online conversations during the conflict,\footnote{\url{https://www.reuters.com/business/media-telecom/facebook-owner-meta-will-block-access-russias-rt-sputnik-eu-2022-02-28/}} Europe's sanctions on Russian state-owned outlets\footnote{\url{https://www.consilium.europa.eu/en/press/press-releases/2022/03/02/eu-imposes-sanctions-on-state-owned-outlets-rt-russia-today-and-sputnik-s-broadcasting-in-the-eu/}} and Russia's\footnote{\url{https://www.theguardian.com/world/2022/mar/04/russia-completely-blocks-access-to-facebook-and-twitter}} ban of Facebook. In particular, as shown in panel \textbf{c}, the median number of daily posts sharing Russian propaganda significantly decreases to 1/4 of the original prevalence (from 0.62\% to 0.15\%, Mann-Whitney $P < .001$), whereas the median number of daily posts linking to low-credibility sources increases by a factor of 2 (from 0.22\% to 0.42\%, Mann-Whitney $P < .001$). Posts sharing links to high-credibility content increase by 50\% (from 0.59\% to 0.79\%, Mann-Whitney $P < .001$). Interestingly, in the Pre-invasion period, the number of posts sharing Russian state outlets is comparable to high-credibility news websites (two-way Mann-Whitney $P = 0.26$).

For what concerns interactions generated by Facebook posts sharing different types of content, we observe similar temporal patterns (see panel \textbf{b} of Figure \ref{fig:prevalence-fb}): a significant increasing trend followed by a decreasing trend for posts linking to Russian propaganda and low-credibility news, and an increasing trend followed by a stationary trend for posts linking to high-credibility content (Mann-Kendall $P< .001$ in all cases). As shown in panel \textbf{d}, the median number of interactions generated by Russian propaganda decreases by over 13 times (from 0.81\% to 0.06\%, Mann-Whitney $P < .001$), whereas low-credibility posts increase their interactions by 3 times (from 0.30\% to 0.95\%, Mann-Whitney $P < .001$); finally, interactions around high-credibility news almost double up (from 1.24\% to 2.20\%, Mann-Whitney $P < .001$). Differently from the number of posts (panel \textbf{c}), in the Pre-invasion period, the median number of interactions generated by reliable sources is higher than Russian propaganda (two-way Mann-Whitney $P < .05$). 

\begin{figure}[!t]
    \centering
    \includegraphics[width=\linewidth]{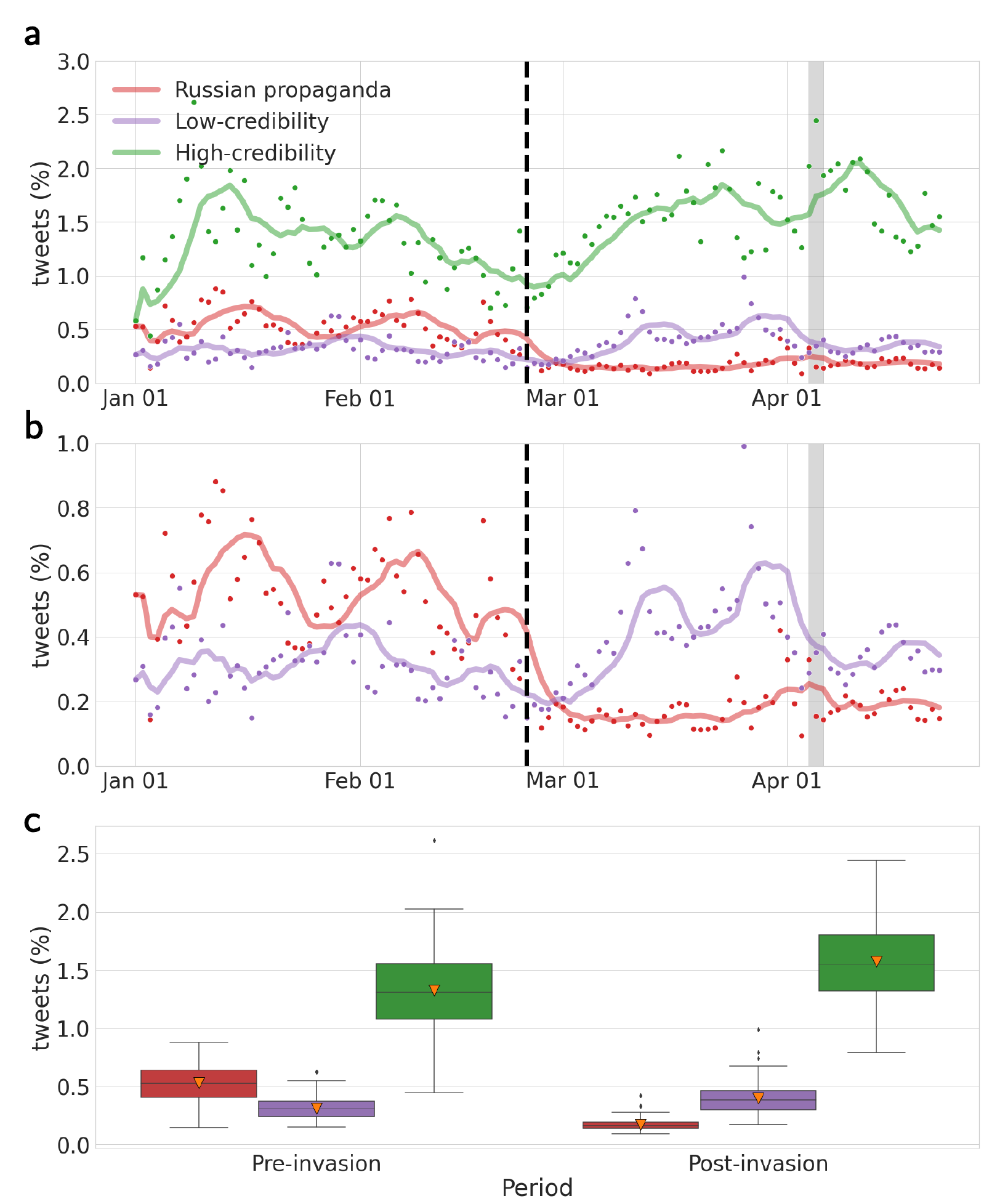}
    \caption{Time series (a,b) and distributions (c) of the daily proportion of tweets sharing links to Russian propaganda, Low-credibility, and High-credibility news websites. Panel (b) provides a zoom-in on Russian and low-credibility sources. The dashed vertical line indicates the day of the invasion (February 24, 2022), and the grey shade indicates a data collection failure. Median values in panel (c) are: Russian propaganda (Pre-invasion = 0.52\%, Post-invasion = 0.16\%); Low-credibility (Pre-invasion = 0.30\%, Post-invasion = 0.38\%); High-credibility (Pre-invasion = 1.30\%, Post-invasion = 1.55\%). Triangles in panel (c) represent the mean value of the distribution.}
    \label{fig:prevalence-tw}
\end{figure}

On Twitter, we identified 567,587 tweets (0.2\% of all tweets) sharing links to Russian propaganda outlets, 997,886 tweets sharing links to low-credibility news websites (0.4\% of all tweets) and 3,949,774 tweets sharing links to high-credibility news websites (1.6\% of all tweets). As shown in panel \textbf{a} of Figure \ref{fig:prevalence-tw}, the number of tweets linking to Russian 
propaganda and low-credibility news does not exhibit a trend in the Pre-invasion period (Mann-Kendall $P > 0.05$); we observe oscillations in the first two months of 2022 (see panel \textbf{b} for a zoom-in of these time series). Instead, tweets sharing high-credibility content exhibit a slightly decreasing trend (Mann-Kendall $P < .05$) before February 24, 2022. Afterward, we observe a significant increasing trend (Mann-Kendall $P < .05$) for tweets linking to low-credibility news (see the two peaks of reshares in March and April) and high-credibility websites, whereas tweets sharing Russian propaganda are stationary (Mann-Kendall $P > 0.05$). As shown in panel \textbf{c}, Russian propaganda is shared more than low-credibility content before the invasion (two-way Mann-Whitney $P < .001$), but after February 24, 2022 the situation reverses (Mann-Kendall $P < .001$), due to Twitter's aggressive policies toward Russian-state outlets\footnote{\url{https://blog.twitter.com/en_us/topics/company/2022/our-ongoing-approach-to-the-war-in-ukraine}} along with Russian ban on the platform and European sanctions on these websites. Specifically, the median number of daily tweets sharing Russian propaganda decreases by over 3 times (from 0.52\% to 0.16\%, Mann-Whitney $P < .001$), and tweets linking to low-credibility sources increase only slightly (from 0.30\% to 0.38\%, Mann-Whitney $P < .05$). Overall, high-credibility outlets are shared significantly more than unreliable websites over the entire period of analysis (two-way Mann-Whitney $P < .001$), and the median daily number of tweets linking to these sources increases only slightly (from 1.30\% to 1.55\%, Mann-Whitney $P < .05$).

\textbf{Findings and remarks:} We assessed the daily prevalence of Russian propaganda and low-credibility content and compared it to a representative sample of reputable and trustworthy sources. We find that Russian propaganda was shared more than generic low-credibility news in the Pre-invasion period, but the relationship reverses in the Post-invasion period, as a consequence of platforms' intervention and European regulations against Russian propaganda, and Russia's ban on Facebook and Twitter. 
Despite content moderation and other intervention policies introduced during the conflict, Russian propaganda does not disappear completely, and misinformation is still present on both platforms in a non-negligible amount. 
On the bright side, overall, high-credibility content was shared more than unreliable sources. 

\subsection{Superspreaders of misinformation}

\begin{figure}
    \centering
\includegraphics[width=\linewidth]{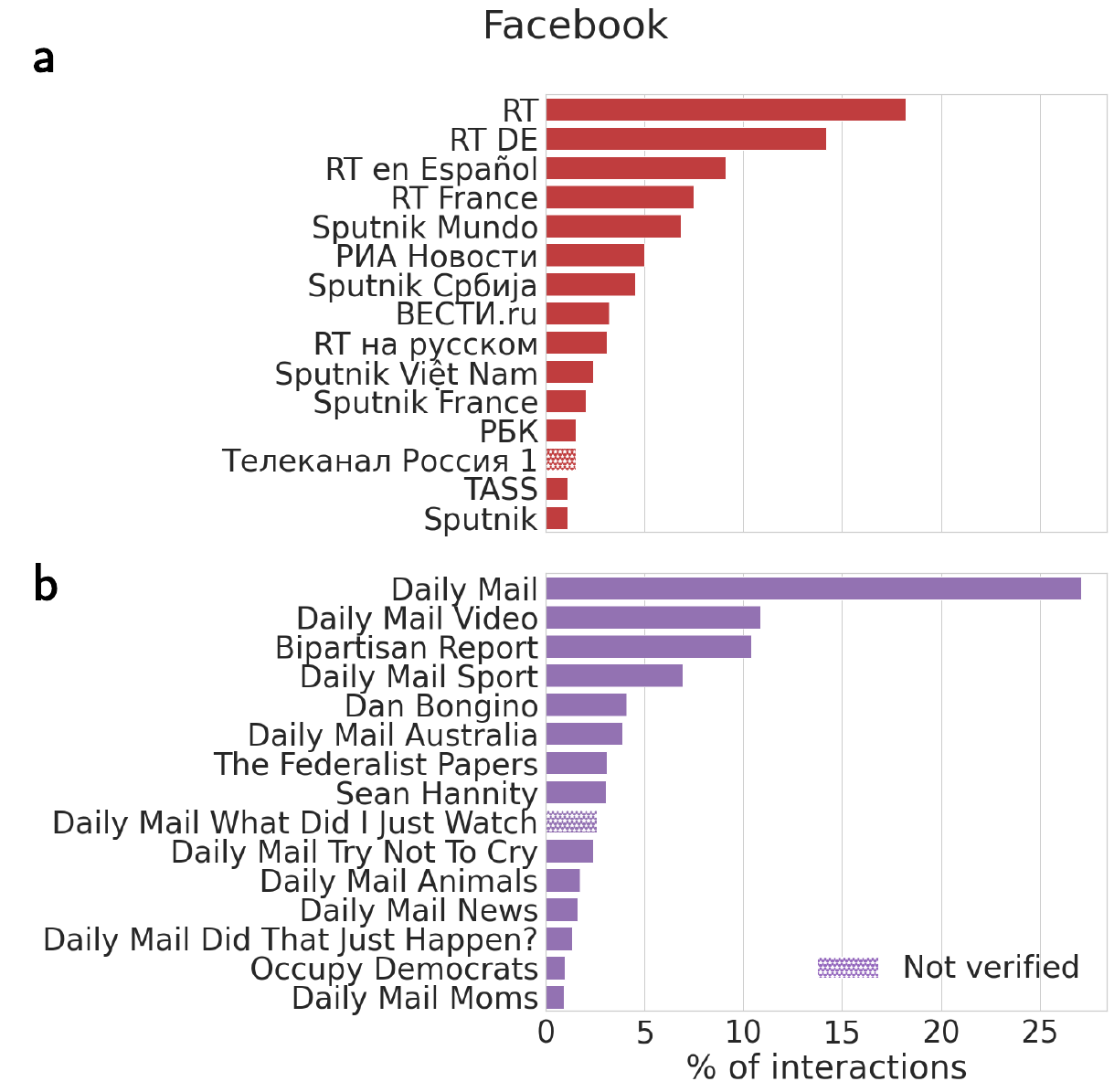}
    \caption{Top 15 spreaders of Russian propaganda \textbf{(a)} and low-credibility content \textbf{(b)} ranked by the proportion of interactions generated over the period of observation, with respect to all interactions around links to websites in each group. Given the large number of verified accounts, we indicate those not verified using ``hatched'' bars.}
    \label{fig:fb-super}
\end{figure}

\begin{figure}
    \centering
    \includegraphics[width=\linewidth]{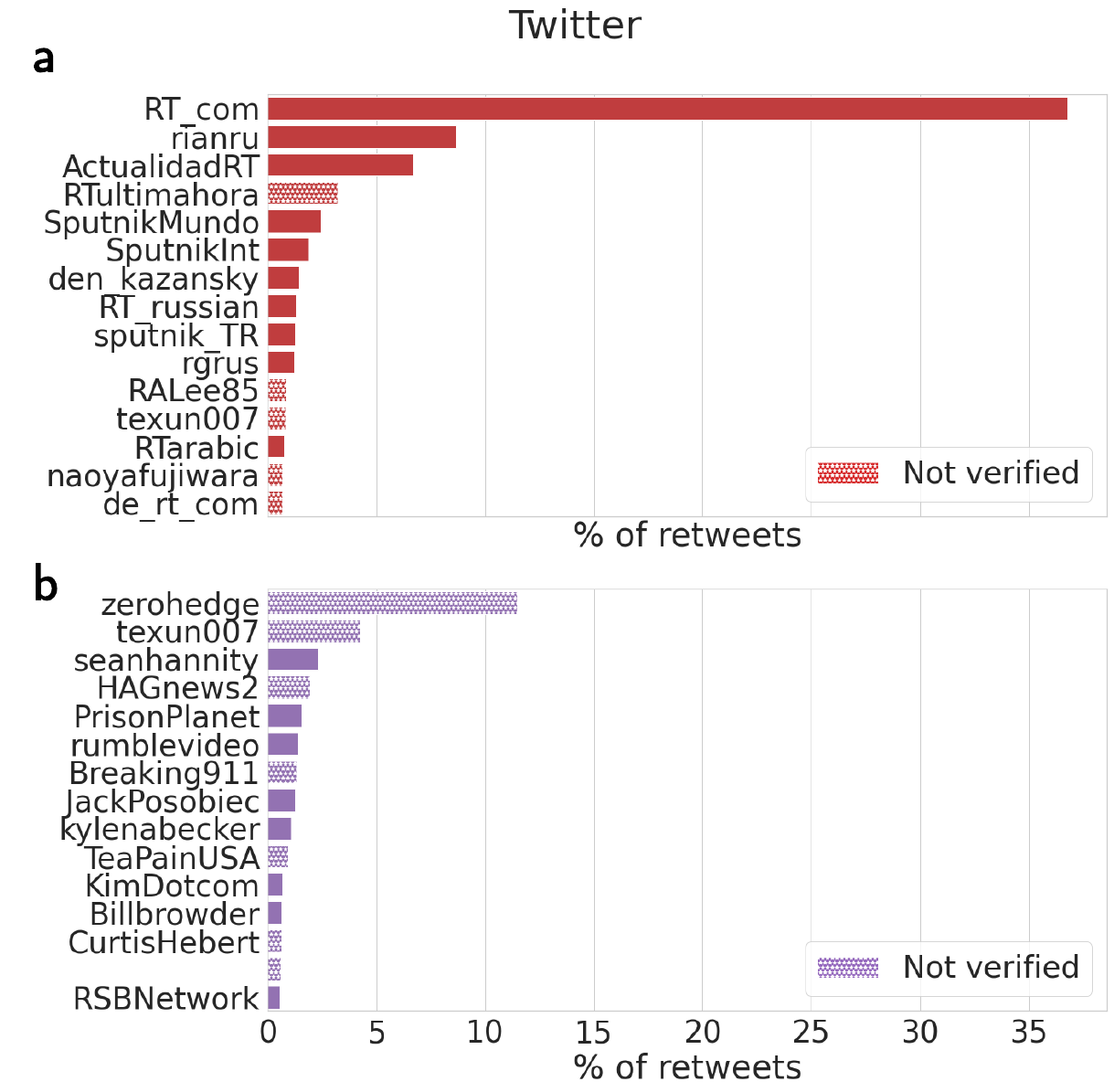}
    \caption{Top 15 spreaders of Russian propaganda \textbf{(a)} and low-credibility content \textbf{(b)} ranked by the proportion of retweets generated over the period of observation, with respect to all retweets linking to websites in each group. We indicate those that are not verified using ``hatched'' bars.}
    \label{fig:tw-super}
\end{figure}

As recent studies suggest that some users might play an outsize role in disseminating misinformation~\cite{nogara2022disinformation, yang2021covid, deverna2022superspreaders, chang2022comparative}, we analyze the role of so-called ``superspreader'' accounts on both Facebook and Twitter with a particular focus on verified accounts. On Twitter, these are accounts ``authentic, notable, and active'' that belong to government, news, entertainment, or another designated category; our period of analysis is prior to Musk's acquisition of the platform,\footnote{\url{https://en.wikipedia.org/wiki/Acquisition_of_Twitter_by_Elon_Musk}} thus we are not considering accounts that subscribe to the new Blue service.\footnote{\url{https://help.twitter.com/en/managing-your-account/about-twitter-verified-accounts}} On Facebook, a verified badge appears when the platform has confirmed that ``the Page or profile is the authentic presence of the public figure or global brand that it represents''.\footnote{\url{https://www.facebook.com/help/196050490547892}}

As shown in Figure \ref{fig:fb-super}, superspreaders of misinformation about the war on Facebook are all verified accounts, with one exception both in the case of Russian Propaganda and generic low-credibility news. We notice mostly accounts associated with outlets and websites (e.g. \texttt{RT} and \texttt{Sputnik Mundo}, \texttt{Daily Mail} and \texttt{Bipartisan Report}, and a few notable right-wing controversial public figures such as Dan Bongino and Sean Hannity. The contribution of the top 15 accounts is disproportionate, as they generate over 80\% of all interactions around links to Russian propaganda and low-credibility information websites.

On Twitter, the picture is very similar in the case of Russian propaganda, where all accounts are verified (with a few exceptions) and mostly associated with news outlets, and generate over 68\% of all retweets linking to these websites (see panel \textbf{a} of Figure \ref{fig:fb-super}). For what concerns low-credibility news, there are both verified (we can notice the presence of \texttt{seanhannity}) and not verified users, and only a few of them are directly associated with websites (e.g. \texttt{zerohedge} or \texttt{Breaking911}). Here the top 15 accounts generate roughly 30\% of all retweets linking to low-credibility websites.

\begin{figure}[!t]
    \centering
    \includegraphics[width=\linewidth]{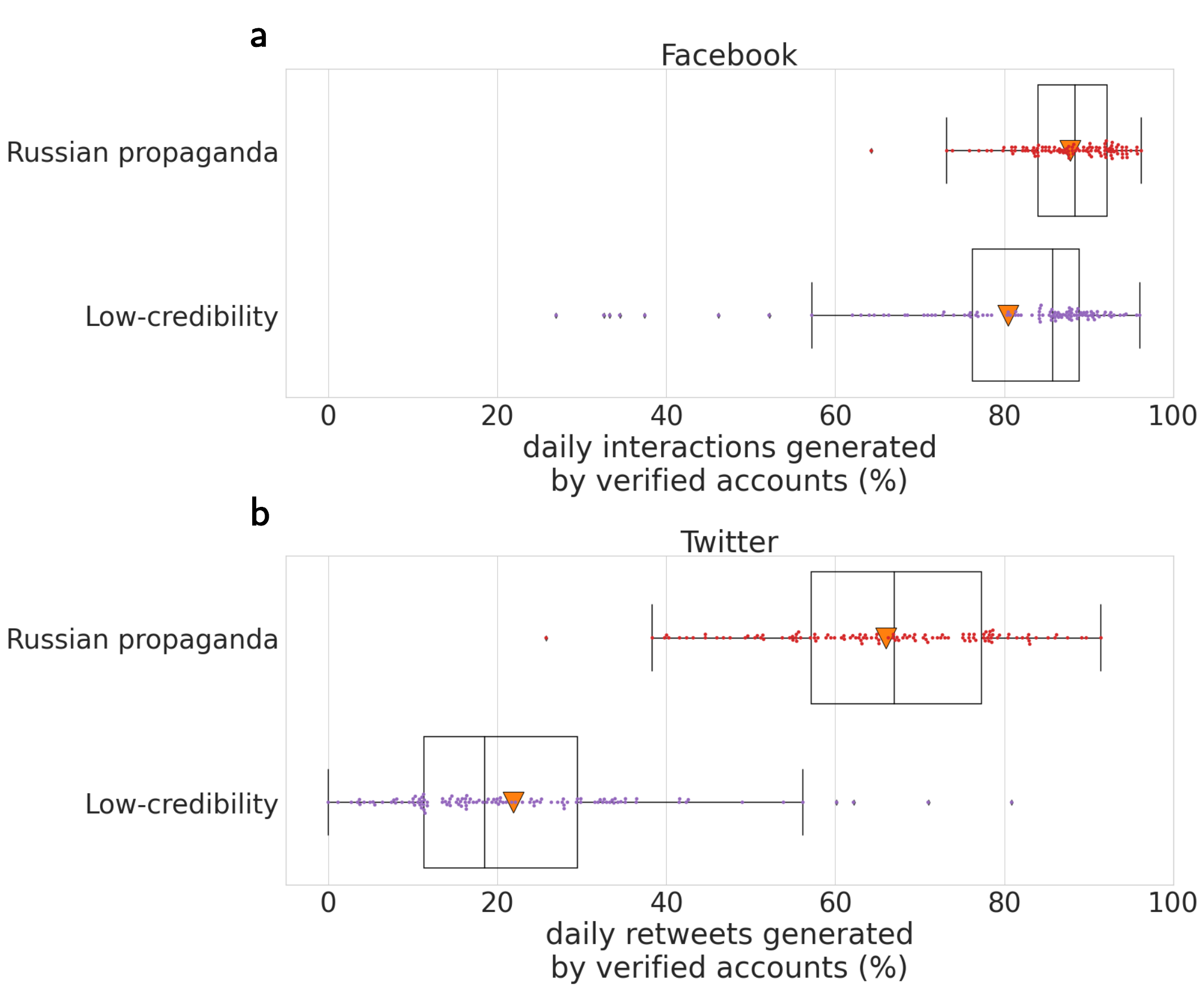}
    \caption{Daily proportion of Facebook interactions (a) and retweets (b) received by posts shared by verified accounts that link to Russian propaganda and low-credibility news. Median values in (a) are: Russian propaganda = 88.34\%, Low-credibility = 85.73\%. Median values in (b) are: Russian propaganda = 66.99\%, Low-credibility = 18.53\%. Triangles represent the mean value of the distribution.}
    \label{fig:verified-daily}
\end{figure}

From a temporal perspective, Figure \ref{fig:verified-daily} shows the daily proportion of Facebook interactions and retweets generated by posts linking to unreliable sources and originally shared by verified accounts. We can see that on an average day over 85\% of the Facebook interactions around links to Russian propaganda and low-credibility news is generated by verified accounts, whereas on Twitter verified accounts contribute for $\sim$67\% of the retweets of Russian propaganda, and 18.5\% for low-credibility news.

\textbf{Findings and remarks:} We estimated the contribution of verified accounts to sharing and amplifying links to Russian propaganda and low-credibility sources, noticing that they have a disproportionate role. In particular, superspreaders of Russian propaganda are mostly accounts verified by both Facebook and Twitter, likely due to Russian state-run outlets having associated accounts with verified status. In the case of generic low-credibility sources, a similar result applies to Facebook but not to Twitter, where we also notice a few superspreaders accounts that are not verified by the platform.  

\subsection{Political leaning of accounts sharing misinformation}
Here, we investigate whether accounts sharing Russian propaganda and low-credibility news lean toward a specific end of the political spectrum. To exclude accounts that were sporadically active on the platforms, we consider only accounts that shared at least 10 posts linking to a website with an assigned political score. Similarly, we build two classes of ``Russian propaganda spreaders'' and ``Low-credibility news spreaders'' by considering only accounts that shared at least 10 posts linking to websites in the corresponding list.\footnote{Results are robust when considering a threshold of 5 posts for political and misinformation posts.} We compare these two classes against all other accounts.

\begin{figure}
    \centering
    \includegraphics[width=\linewidth]{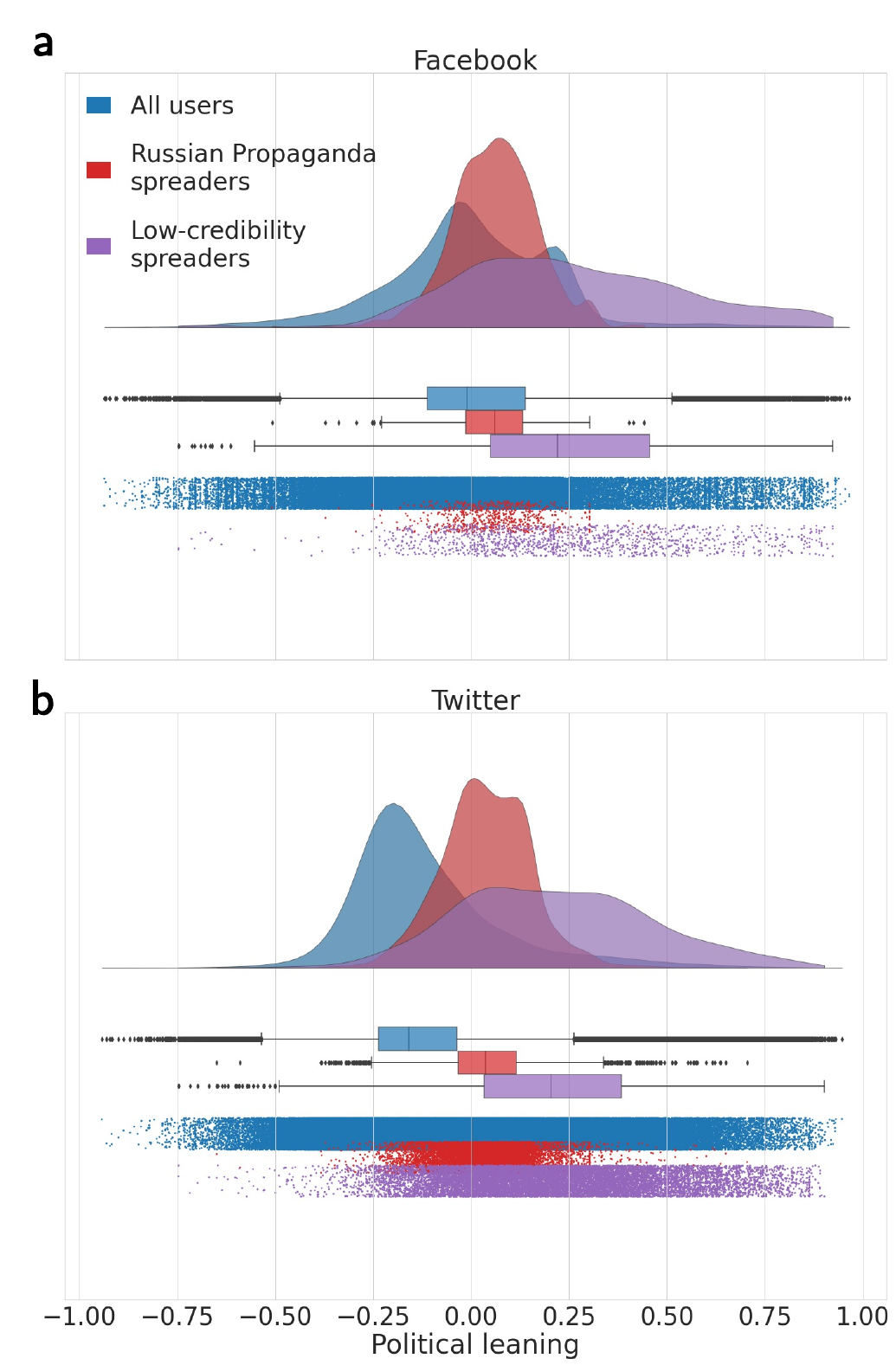}
    \caption{Distribution of political leaning score for accounts sharing Russian propaganda and low-credibility content, and all other users, respectively for Facebook \textbf{(a)} and Twitter \textbf{(b)}. Median values in (a) are: Russian propaganda spreaders = $0.03$, Low-credibility spreaders  = $0.20$, All users = $-0.15$. Median values in (b) are: Russian propaganda spreaders = $0.03$, Low-credibility spreaders = $0.20$, All users = $-0.15$}
    \label{fig:political}
\end{figure}

As shown in panel \textbf{a} of Figure \ref{fig:political}, Facebook accounts sharing links to misinformation (either Russian propaganda or low-credibility) websites are more right-leaning than the average account. These results, which are significant according to two-way Mann-Whitney tests ($P < .001$ in each pairwise test), are in accordance with existing literature on the interplay between political leaning and the spread of misinformation \cite{grinbergFakeNewsTwitter2019,nikolovRightLeftPartisanship2021}.

Panel \textbf{b} of Figure \ref{fig:political} shows that, also on Twitter, accounts sharing links to unreliable sources are more right-leaning than the average account, confirming previous findings from the literature. Similar to Facebook, the political leaning of accounts sharing low-credibility news websites is skewed much more towards the right compared to Russian propaganda spreaders. An application of two-way Mann-Whitney tests confirms the significance of these results ($P < .001$ in each pairwise test).

\textbf{Findings and remarks:} We inferred the political leaning of accounts based on the number of political URLs they shared, and compared those being active at spreading Russian propaganda and low-credibility news versus the others. We found that, in accordance with existing literature, accounts sharing misinformation tend to be more right-leaning than the average account, and that this discrepancy is more accentuated for generic low-credibility news than Russian propaganda.

\subsection{Content moderation by platforms}
We finally analyze the efforts of Facebook and Twitter at removing misleading and unsubstantiated information that violated the platforms' terms during the conflict. We remark that we only estimate a sample of posts removed on Facebook due to API limitations, whereas we identified all tweets that were not available as of October 2022.

\begin{figure}[!t]
    \centering
    \includegraphics[width=\linewidth]{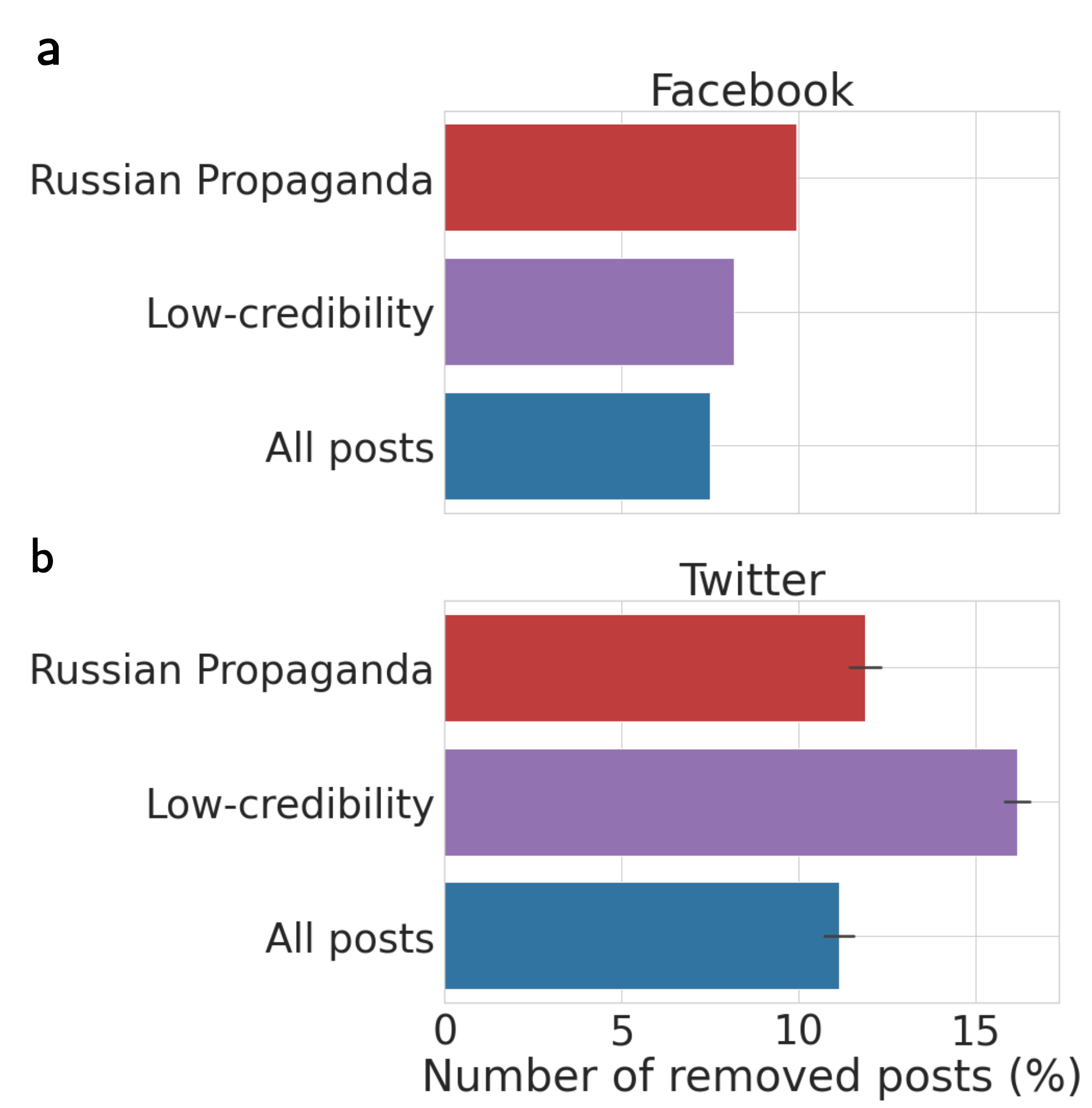}
    \caption{Proportion of posts that are not accessible on Facebook \textbf{(a)} and Twitter \textbf{(b)}, among those sharing links to Russian propaganda outlets, low-credibility, and all posts. For Facebook, we estimate the proportion of removed posts on samples of 10K posts for each category. For Twitter, we show the distribution of daily proportions of tweets removed among those linking to different groups of websites. Error bars in (b) represent the standard error on the mean daily value.}
    \label{fig:removed}
\end{figure}

Figure \ref{fig:removed} shows the proportion of removed posts among those sharing links to Russian propaganda outlets and low-credibility news websites. We also consider posts that were removed regardless of whether they shared links to information sources or not.

On both platforms, the proportion of removed posts linking to unreliable websites is higher than a random post. On Facebook, we estimated that $\sim10\%$ of posts linking to Russian propaganda, and over $8\%$ of posts linking to low-credibility news were removed on average, compared to $\sim7\%$ of random posts. On Twitter, we found that $\sim11.9\%$ of tweets linking to Russian propaganda, and over $15\%$ of tweets linking to low-credibility news were removed on average, compared to $\sim11\%$ of posts linking to high-credibility information websites. We also observe that, while tweets sharing links to low-credibility news are more likely to be removed than those linking to Russian propaganda websites, the converse applies on Facebook, where posts sharing Russian propaganda were more likely to be removed.

\textbf{Findings and remarks:} We estimated the proportion of misleading content that is not present on platforms anymore, either because Facebook or Twitter removed the content or deactivated their author, or because accounts deliberately removed it. We found that posts sharing links to misinformation sources were more likely to disappear from the platform, although not completely, compared to random posts related to the conflict. Overall, only 8-15\% of posts sharing links to unreliable sources were removed by both platforms.

\section{Conclusions}
\subsection{Contributions and Findings}
We carried out a longitudinal study of the spread of misinformation about the Russian invasion of Ukraine, originating from Russian state outlets and low-credibility sources shared on Twitter and Facebook during the first months of the conflict. We highlighted a considerable drop in the prevalence of Russian propaganda following the invasion, as a consequence of new platforms' policies, European regulations of Russian propaganda and the Russian ban on online social networks. Throughout the period of analysis, misinformation is generally less prevalent and generates fewer interactions than high-credibility content, but its presence is not negligible. We showed that a few accounts yield a disproportionate role in spreading and amplifying misinformation, and in most cases, they have a verified badge on the platforms. We estimated that accounts sharing misinformation exhibit a right-wing political leaning compared to the general sample of accounts discussing the war on both platforms. Finally, we measured the amount of misinformation content that was removed by the platforms, showing that posts sharing links to Russian propaganda outlets and low-credibility sources are in general more likely to be removed than random posts. However, misinformation does not disappear completely from the platforms as only 8-15\% of Facebook posts and tweets linking to Russian propaganda and low-credibility news websites were removed.

\subsection{Limitations}
Our study does not come without limitations. Collecting data from Twitter is hindered by the 1\% limit on the streaming API, which might have biased our collection in the first weeks of the invasion \cite{morstatter2013sample}. When identifying links to misinformation websites, we did not consider shortened links -- e.g., web domains such as \url{bit.ly} that are often expanded by platforms themselves -- and, therefore, our estimates of Russian propaganda, low-credibility, and high-credibility could be lower than the actual numbers \cite{yang2021covid}. Besides, we only consider misinformation shared via news articles, thus ignoring what might come from multimedia content such as photos, videos and memes. We did not manually verify the articles published from misinformation sources, but relied on literature supporting the distant-supervision approach to identify misinformation at scale \cite{Lazer-fake-news-2018}. Also, our approach to assess the amount of content removed by the platforms is not perfect, and it does not allow us to ascertain the exact reasons for the removal. In all our analyses, we did not account for the activity of automated accounts, which might play a role in the spread of misinformation \cite{shao2018spread}. Finally, our data mostly captures conversations on Western-centric platforms, thus overlooking countries such as Ukraine and Russia, and bth Facebook and Twitter exhibit demographic biases in their user base, which are not completely representative of the general population \cite{pew2019social}.

\subsection{Discussion and Future Work}
There are several implications of our findings. First and foremost, having access to reliable and accurate online information during crises is crucial to preserve the democratic process. Our results show that, while the prevalence of high-credibility news articles was generally higher than misinformation, the latter was still present on the platform throughout the period of analysis (including platforms' ban on Russian propaganda), generating over 65 M interactions on Facebook and 1 M retweets. This indicates that platforms' efforts to preserve the integrity of online conversations were not successful enough. Second, we highlighted the role of certain influential accounts, so-called ``superspreaders'' of misinformation, in promoting and amplifying misinformation. A handful of them was responsible for 60-80\% of all interactions and retweets of misinformation, and they were mostly verified by the platforms. Research suggests that they are often driven by financial incentives \cite{deverna2022superspreaders,pierri2022one,nogara2022disinformation}, and platforms might consider several strategies to reduce their impact, including revoking their verified status, down-ranking their content, or making it not visible to users (``shadowbanning''). Finally, our results provide evidence of similar patterns in the landscape of misinformation across two different platforms such as Facebook and Twitter, contributing to the existing literature that focuses on cross-platform analyses of cyber-social threats. 

Future work could consider similar analyses on other (niche) platforms (e.g. Gab, Parler, 4chan, etc.) where misinformation can originate before migrating toward more mainstream media. The content and spreading patterns of different kinds of misinformation, such as photos, videos, and memes should be studied. Researchers might be interested in monitoring the spread of ``domestic'' propaganda that originates from hyper-partisan pundits and political figures. Finally, future research might also aim to estimate the real-world consequences of the spread of misinformation related to the conflict, especially in the context of the 2022 U.S. Midterm election.

\subsection{Ethical Concerns}
In our analyses, we do not attempt to identify or de-anonymize individual users nor do we share their inferred political leaning -- with the exception of a handful of superspreaders of misinformation, most of which are public figures verified by platforms. We present data collected through public APIs in an aggregated fashion, and we transparently provide access to the IDs of Facebook posts and tweets. These can be used to retrieve the dataset in accordance with platforms' data-sharing policies, with the exception of posts that have been removed or made private by users, thus limiting reproducible analyses. We also provide access to auxiliary material in order to replicate our findings. Our study is observational and retrospective, thus users were not harmed in the process. The project was approved by our IRB (\#UP-21-00005-AM001).

\section{Acknowledgements}
Work supported in part by
DARPA (contract \#HR001121C0169) and PRIN grant HOPE (FP6, Italian Ministry of Education). We are thankful to Emily Chen for kindly providing access to Twitter data. Any opinions, findings, and conclusions or recommendations expressed in this paper are those of the authors and do not necessarily reflect the views of the funding agencies.

\bibliographystyle{ACM-Reference-Format}
\bibliography{bib}

\end{document}